\documentclass[aps,prb,twocolumn,superscriptaddress,raggedbottom,showpacs,floatfix]{revtex4-1}
\usepackage{color}
\usepackage{graphicx,latexsym}
\usepackage{epstopdf}
\usepackage{comment,bm}

\usepackage{enumerate}
\usepackage{graphicx}
\usepackage{amsmath}
\usepackage{amssymb}
\usepackage{mathtools}
\usepackage{graphics}
\usepackage{graphicx}
\usepackage{comment}
\usepackage{tikz}
\usepackage{mathtools}
\usepackage{bm}

\newcommand{\beq}{\begin{equation}}
\newcommand{\eeq}{\end{equation}}
\newcommand{\bei}{\begin{itemize}}
\newcommand{\eei}{\end{itemize}}
\newcommand{\ben}{\begin{enumerate}}
\newcommand{\een}{\end{enumerate}}

\newcommand{\bB}{{\mathbf B}}

\newcommand{\be}{{\mathbf e}}
\newcommand{\bE}{{\mathbf E}}
\newcommand{\bj}{{\mathbf j}}

\newcommand{\bS}{{\mathbf S}}
\newcommand{\bp}{{\mathbf p}}

\newcommand{\bq}{{\mathbf q}}
\newcommand{\br}{{\mathbf r}}
\newcommand{\bu}{{\mathbf u}}

\newcommand{\bk}{{\mathbf k}}
\newcommand{\bw}{{\mathbf w}}
\newcommand{\bnabla}{{\boldsymbol\nabla}}

\newcommand{\vecf}{\mathbf f}

\newcommand{\ket}[1]{|#1\rangle}
\newcommand{\bra}[1]{\langle #1|}

\usepackage{hyperref}
\hypersetup{
   pdfpagemode=None, 
   pdfstartpage=1,
   pdfmenubar=true,
   pdftoolbar=true,
   colorlinks = true,
   linkcolor=blue,
   citecolor=blue,
   urlcolor=blue,
   bookmarksopen=false
 }

\definecolor{darkblue}{rgb}{0.,0.24,0.51}
\definecolor{britishracinggreen}{rgb}{0.0, 0.26, 0.15}
\definecolor{darkgreen}{rgb}{0,0.60,.2}

\def\be{\begin{equation}}
\def\ee{\end{equation}}

\newcommand{\tr}{\textcolor{red}} 

\begin{document}
\renewcommand{\vec}{\mathbf}
\renewcommand{\Re}{\mathop{\mathrm{Re}}\nolimits}
\renewcommand{\Im}{\mathop{\mathrm{Im}}\nolimits}
\newcommand\scalemath[2]{\scalebox{#1}{\mbox{\ensuremath{\displaystyle #2}}}}

\title{Weyl excitations via
helicon-phonon mixing in conducting materials}


\author{Dmitry K. Efimkin}
\email{dmitry.efimkin@monash.edu}
\affiliation{School of Physics and Astronomy, Monash University, Victoria 3800, Australia}
\affiliation{ARC Centre of Excellence in Future Low-Energy Electronics Technologies, Monash University, Victoria 3800, Australia}

\author{Sergey Syzranov}
\affiliation{Physics Department, University of California, Santa Cruz, California 95064, USA}

\begin{abstract}
Quasiparticles with Weyl dispersion can display an abundance of novel topological, thermodynamic and transport phenomena, which is why 
novel Weyl materials and platforms for Weyl physics are being intensively looked for in electronic, magnetic, photonic and acoustic systems.
We demonstrate that conducting materials in magnetic fields generically host Weyl excitations due to the hybridisation of phonons with helicons, collective neutral modes of electrons interacting with electromagnetic waves propagating in the material.
Such Weyl excitations are, in general, created by the interactions of helicons with longitudinal acoustic phonons. An additional type of Weyl excitation in polar crystals comes from the interaction between helicons and longitudinal optical phonons.
Such excitations can be detected in X-ray and Raman scattering experiments. 
The existence of the Weyl excitations involving optical phonons in the bulk of the materials also leads to the formation of topologically protected surface arc states that can be detected via surface plasmon resonance.
\end{abstract}

\maketitle

\textcolor{blue}{\emph{Introduction}}. The tremendous recent interest 
in Weyl materials~\cite{WanVishwanath:WeylFirst,Armitage:WeylReview,Xu:TaAs, Lv:discovery, Lv:2015, Xu:WeylDiscovery, Xu:NbAs, Yang:TaAs, Xu:TaP,  ZHasan:reviewDiscovery} is owed, in large part, to 
a plethora of fundamental and novel 
topological phenomena they can display: the chiral anomaly~\cite{Burkov:IOPcmeReview, Burkov:Annals2018}, topological surface states~\cite{WanVishwanath:WeylFirst,Armitage:WeylReview}, unconventional regimes of transport~\cite{SonSpivak:NMR,SyzranovRadzihovsky:review,Moll:SlabOscillations,Potter:OscillationsPrediction,Skinner:puddles}, etc.
Numerous predictions and observations of Weyl-related topological phenomena have motivated researchers to look for Weyl excitations not only in electronic but also in magnetic, photonic and acoustic systems.





Excitations with Weyl dispersion may generically be engineered via the hybridisation of two other types of excitations with similar energies if the interactions between them vanishes along a certain direction of momentum.
Such interactions lead to a degeneracy between the two bands of the hybridised excitations at the respective wave vectors and a gap for other wave vectors.
For example, the hybridization between different types of plasma waves~\cite{HeliconReview1,HeliconReview2} has been shown to lead to Weyl excitations
in photonic metamaterials~\cite{WeylMM1,WeylMM2,WeylMM3} and magnetized electron gases~\cite{PlasmaTop1,PlasmaTop2,PlasmaTop3,PlasmaTop4}. This mechanism of the formation of Weyl excitations is similar to the emergence of 2D topological excitations due to the hybridization between phonons and spin-waves~\cite{
MagnonPhonon1,MagnonPhonon2,MagnonPhonon3,MagnonPhonon4,MagnonPhonon5,MagnonPhonon6,MagnonPhonon7,MagnonPhonon8}, spin- and plasma-waves~\cite{EfimkinSpinPlasma}, as well as excitons and cavity photons~\cite{TopPolaritons1,TopPolaritons2}.


		

In this Letter, we demonstrate that conducting materials in magnetic fields generically host Weyl excitation due to the hybridisation of phonons
with helicons~\cite{HeliconReview1,HeliconReview2}, collective neutral modes of electrons interacting with electromagnetic waves propagating in the material.
Such Weyl excitations emerge generically due to the hybridisation of helicons with {longitudinal} acoustic phonons. 
In polar crystals, Weyl excitations are also created by the interactions of helicons with {longitudinal} optical phonons.

The predicted Weyl excitations can be detected in, e.g., Raman-spectroscopy experiments. 
Weyl excitations involving optical phonons in the bulk of the material also lead to the formation of topologically protected arc states on its surfaces, which can be observed {via the} surface plasmon resonance.

\begin{figure*}[t]
\begin{center}
    \includegraphics[width=\textwidth]{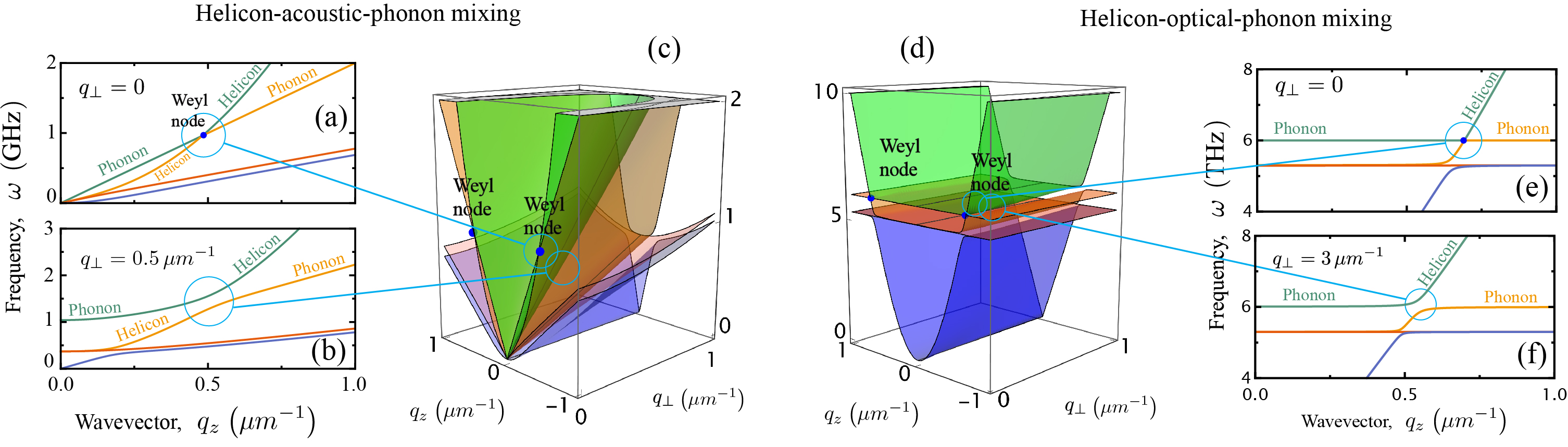}
   \caption{\label{fig:AllDispersions}
   The dispersion of hybrid
   modes emerging due to the hybridisation of plasma waves with acoustic (a-c) and optical (d-f) phonons in a conductor in a magnetic field.
   The 3D plots (c) and (d)
   shows the dispersion for as a function of the wave vectors $q_z$ and $q_\bot$ along and perpendicular to the magnetic field. 
   The side panels are 2D plots of the dispersion as a function of $q_z$ for different values of $q_\bot$. As a result of the hybridisation of helicon and longitudinal-phonon branches, Weyl points emerge at $q_\bot=0$, as shown in panels (a) and (c) for the case of acoustic phonons and in panels (d) and (e) for the case of optical phonons.
   The numbers on the axes are given for a potassium crystal in the magnetic field $B=20\; \hbox{T}$ (a-c)
   and for heavily doped~\cite{SM} $\hbox{CdSe}$ with the electron density $n=4\cdot 10^{19}~\mathrm{cm}^{-3}$ in the magnetic field $B=45\; \hbox{T}$. (d-f)} 
\end{center}
\end{figure*}

\textcolor{blue}{\emph{Heuristic argument for the emergence of Weyl nodes}}. 
A conductor in a magnetic field hosts excitations called {\it helicons}~\cite{HeliconReview1} that exhibit the anisotropic dispersion $\propto |\bq|(\bB\cdot \bq)$ at small wave vectors $\bq$. 
If helicon-phonon interactions are neglected, the helicon dispersion intersects along a line with the phonon dispersion. 
Such an intersection always exists for acoustic phonons (see Fig.~\ref{fig:AllDispersions}c), which are linearly dispersive for small $\bq$. It also exists for optical phonons, which are gapped, if the magnetic field (cyclotron frequency $\omega_c$) is sufficiently large (see Fig.~\ref{fig:AllDispersions}d).

Unless wavevector $\vec{q}$ is aligned with the direction of the magnetic field $\vec{B}$,  helicons involve both transverse and longitudinal oscillations of electric current and electromagnetic fields. As a result, they interact with both longitudinal and transverse phonon modes and the  mixed helicon-phonon excitations are separated by a gap (as shown in Figs.~\ref{fig:AllDispersions}b and f). In contrast, helicons propagating along the magnetic field $\vec{B}$ are purely transverse circularly polarized waves (these corkscrew-shaped oscillations motivated the term ``helicon") and their coupling with the longitudinal phonon mode vanishes (cf. Figs.~\ref{fig:AllDispersions}a and e). As a result, dispersion relations for the mixed helicon-phonon excitations intersect at two selected points in reciprocal space and are of the Weyl nature in their vicinity. 

The intersections of the dispersion relations for the mixed modes involving acoustic phonons and helicons have been known previously~\cite{HeliconReview1}. In this Letter, we demonstrate that they are of the Weyl nature and that such excitations also emerge for optical phonons. Furthermore, we show that Weyl excitations created by optical-phonon-Weyl hybridisation can be readily observed in experiments and favor the emergence of arc surface modes.




\emph{\color{blue} Helicon waves.}
In the absence of phonons, the propagation of the helicon waves is described by the linearized hydrodynamics equation
\begin{subequations}
\begin{align}
\label{Continuity}
\partial_t \vec{j}(\vec{r},t) &=\frac{ne^2}{m} \vec{E}(\vec{r},t) - \frac{e}{mc}[\vec{j}(\vec{r},t)\times \vec{B}], 
\end{align}
complemented by the Maxwell equations
\begin{align}
\label{Maxwell}
\mathrm{curl}\, \vec{H}(\vec{r},t) & = \frac{4\pi \vec{j}(\vec{r},t)}{c}+\frac{1}{c} \partial_t \vec{D} (\vec{r},t), \\
\mathrm{curl}\, \vec{E}(\vec{r},t) & = -\frac{1}{c} \partial_t \vec{H} (\vec{r},t),
\label{Maxwell2}
\end{align}
\end{subequations}
where $e$ is the absolute value of the electron charge; $n$ is the electron density; $m$ is the effective mass and $c$ is the speed of light in vacuum. The magnetic field includes both the large external magnetic field $\vec{B}=B \vec{e}_\mathrm{z}$ and a small oscillating component $\vec{H}(\vec{r},t)$ created by propagating current density oscillations. 
For simplicity, we assume that the medium is non-magnetic, i.e. has 
the magnetic permeability $\mu=1$.


The helicon dispersion 
can be obtained from Eqs.~(\ref{Continuity})-(\ref{Maxwell}) as a solution of the eigenvalue problem $\omega \psi_\mathrm{pl}=\hat{\mathcal{H}}_\mathrm{pl}(\vec{q}) \psi_\mathrm{pl}$~\cite{PlasmaTop1,PlasmaTop2,PlasmaTop3,PlasmaTop4} with the $9$-component vector $\psi_\mathrm{pl}=\{\sqrt{\epsilon_{\infty}}\vec{E}(\vec{q}), \vec{H}(\vec{q}), 4\pi\vec{j}(\vec{q})/\omega_{\mathrm{p}} \sqrt{\epsilon_{\infty}}\}^T$, where
$\omega_{\mathrm{p}}=\sqrt{4\pi ne^2/m\varepsilon_\infty}$
is the plasma frequency and 
$\varepsilon_\infty$ is the contribution of the electronic shells of the atoms to
the dielectirc constants \{at GHz and higher frequencies, $\vec{D}(\vec{r},t)=\epsilon_\infty \vec{E}(\vec{r},t)$\}.
The matrix $\hat{\mathcal{H}}_\mathrm{pl}$ is given by
\begin{subequations}
\begin{equation}
\label{HamiltonianPlasma}
\hat{\mathcal{H}}_\mathrm{pl}(\vec{q})=\begin{pmatrix}
0 & - \bar{c} \hat{K}_{\vec{q}} & - i \omega_\mathrm{p} \hat{1}\\
\bar{c} \hat{K}_{\vec{q}} & 0 & 0 \\
 i \omega_\mathrm{p} \hat{1} & 0 & -i \omega_{\mathrm{c}} \hat{L}  \\
\end{pmatrix},
\end{equation}
where we have introduced the speed $\bar{c}=c/\sqrt{\epsilon_\infty}$ 
of electromagnetic waves in the material and the
antisymmetric matricies $\hat{K}_\vec{\vec{q}}$ and $\hat{L}$ are given by 
\begin{equation}
    \hat{K}_{\vec{q}}=\begin{pmatrix}
0 & -q_z  & q_y \\
q_z & 0 & -q_x \\-q_y & q_x & 0  \\ 
\end{pmatrix}, \quad
\hat{L}=\begin{pmatrix}
0 & 1  & 0 \\
-1 & 0 & 0 \\0 & 0 & 0  \\
\end{pmatrix},
\label{Ham2}
\end{equation}
where $\omega_\mathrm{c}=eB/mc$
is the cyclotron frequency.
\end{subequations}

\begin{figure}[b]
	\label{FigDispersions}
	\begin{center}
		\includegraphics[trim=0cm 0cm 0cm 0cm, clip, width=0.85\columnwidth]{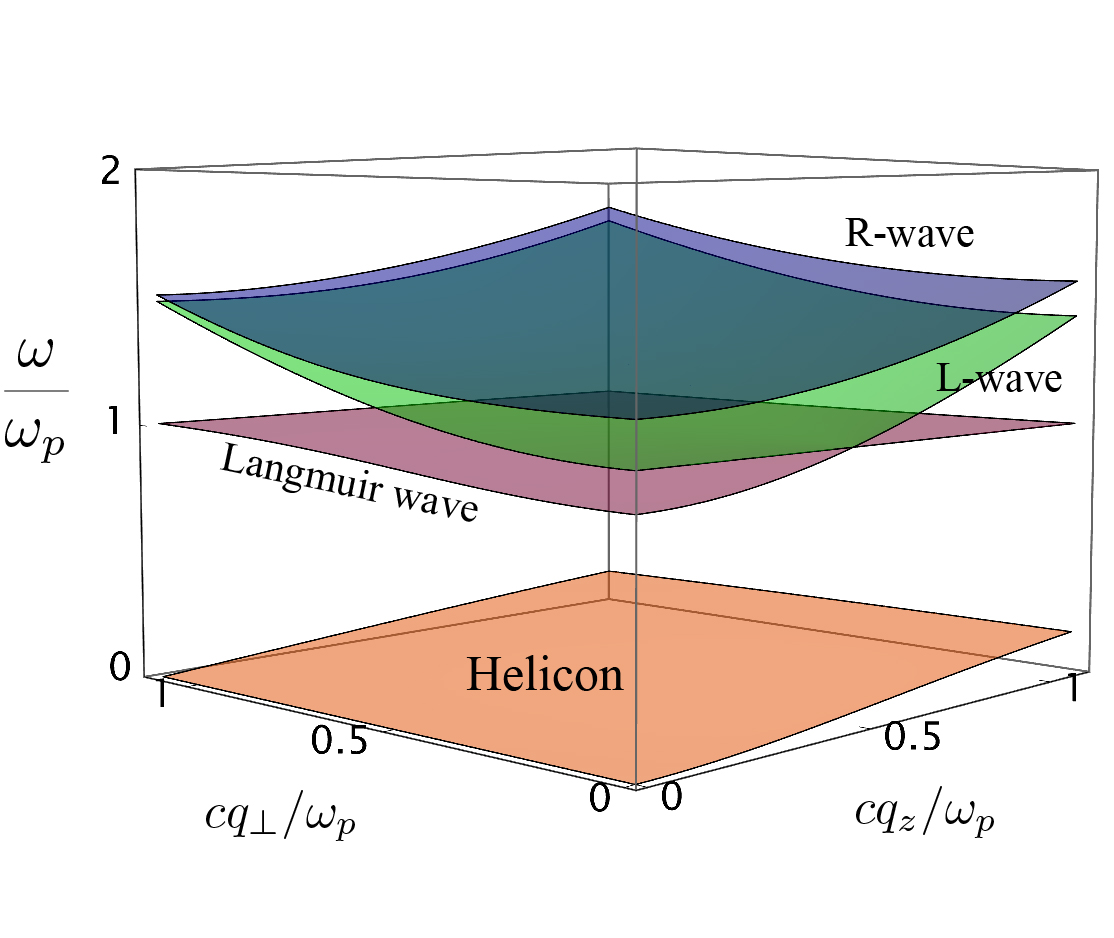}
		\vspace{-0.3in}
		\caption{\label{PlotPlasma}
   The dispersion of plasma waves in an electron gas in a magnetic field for $\omega_\mathrm{c}/\omega_\mathrm{p}=0.3$. They include three high-frequency modes (Langmuir, R-, and L-waves with frequencies around the plasma frequency, $\omega_\mathrm{p}$) and the low-frequency helicon mode, whose dispersion is  anisotropic and bounded by the cyclotron frequency, $\omega_\mathrm{c}$.} 
	\end{center}
	\vspace{-0.2in}
\end{figure} 

The dispersion of the magnetoplasma modes is given by 
the positive ($\omega>0$)
eigenvalues of the 
``Hamiltonian''
\eqref{HamiltonianPlasma}.
They include three gapped
dispersion branches (longitudinal Langmuir waves and transverse circularly polarized L- and R-waves) with the gap {of the order of} the plasma frequency $\omega_p$ and a gapless branch, helicons (sometimes referred to as whistlers~\cite{PlasmaBook}, cf. Fig.~\ref{PlotPlasma}), on which we focus in the rest of the paper.

For small ratios $\omega_c/\omega_p$ of the cyclotron to plasma frequencies, corresponding to realistic materials, the high-frequency modes can be excluded (see Supplemental Material~\cite{SM} for details) and the dynamics of helicons can be conveniently described using the three-component vector 
$\psi_\mathrm{hel}=\left\{(B_x-iB_y)/\sqrt{2},-B_z,(-B_x-iB_y)/\sqrt{2}\right\}$. The dispersion $\omega(\vec{q})$ of the low-energy excitations is given by the eigenvalue problem $\omega(\vec{q})\psi_\mathrm{hel}=\hat{\mathcal{H}}_{\mathrm{hel}}(\vec{q})\psi_\mathrm{hel}$ with the matrix $\hat{\mathcal{H}}_{\mathrm{hel}}(\vec{q})$
given by
\begin{equation}
\label{HamiltonianH}
\hat{\mathcal{H}}_{\mathrm{hel}}(\vec{q})
=\alpha q_z\begin{pmatrix} q_z & \frac{q_x+i q_y }{\sqrt{2}}  & 0 \\ \frac{q_x-i q_y }{\sqrt{2}} & 0 & \frac{q_x+i q_y }{\sqrt{2}} \\ 0 & \frac{q_x-i q_y}{\sqrt{2}} & -q_z
\end{pmatrix},    
\end{equation}
where $\alpha=\omega_\mathrm{c} \bar{c}^2/\omega_\mathrm{p}^2=
B c/4\pi n e$.

By introducing the effective spin-$1$ operator $\bf\hat s$ in the space of vectors $\psi_\mathrm{H}$, the operator~\eqref{HamiltonianH} can be rewritten in the form $\hat{\mathcal{H}}_{\mathrm{hel}}(\vec{q})=\alpha q_z\left({\bf\hat s}\cdot \vec{q}\right)$ similar to the dispersion $\hat H_\text{Weyl}=v\left({\bf\hat s}\cdot \vec{q}\right)$ of a 
spin-$1$ Weyl semimetal~\cite{BradlynBernevig:WeylSpin1}. 
The topological properties of the low-energy helicons are similar to those of spin-$1$ Weyl fermions.
The operator~\eqref{HamiltonianH} describes the helicon dispersion for small wave vectors $|\vec{q}|\ll \omega_p/c$. At larger wave vectors, the dispersion $\omega(\vec{q})$ with nonvanishing transverse wave vectors saturates to $\omega_c$. 

We note that helicons propagating (anti-)parallel to the magnetic field $\bB$ have circular polarization in the plane perpendicular to $\bB$.
As we discuss below, the transverse polarization of helicons leads to the vanishing of the helicon-phonon interaction for both {longitudinal} acoustic and longitudinal optical phonons. 


\textcolor{blue}{\emph{Interactions between helicons and acoustic phonons.}} In the absence of electrons, the dynamics of
long-wave acoustic phonons is 
universally described in terms of the lattice displacement $\bu(\br,t)$ by the equation of motion~\cite{LL7}
\begin{subequations}
\begin{equation}
\label{PhononsAcoustic1}
\partial_t^2\vec{u}(\br,t)=s_\mathrm{t}^2 \Delta \vec{u}(\br,t) + (s_\mathrm{l}^2-s_\mathrm{t}^2) \nabla [\nabla \cdot \vec{u}(\br,t)]
\end{equation}
or the equivalent Hamilton's equations
\begin{align}
\partial_t\vec{u}(\bq,t)=\vec{p}(\bq,t), 
\quad
\partial_t\vec{p}(\bq,t)=-(\hat{W}_{\vec{q}}^{\mathrm{ac}})^2 \vec{u}(\bq,t),
\end{align}
\end{subequations}
where $\hat{W}^\mathrm{ac}_\bq$, $\vec{u}(\vec{q},t)$ and $\vec{p}(\vec{q},t)$ are the Fourier-transforms of, respectively, 
$\hat{W}^\mathrm{ac}_{\br-\br^\prime}$, $\vec{u}(\vec{r},t)$ and $\vec{p}(\vec{r},t)$. The momentum $\bp$ is the momentum canonically conjugate to the displacement $\bu$; 
$s_{\mathrm{l}}$ and $s_{\mathrm{t}}$ 
are the velocities of, respectively, longitudinal and transverse phonons;
the matrix $\hat{W}^{\mathrm{ac}}_{\vec{q}}$ (dynamical matrix) describes the energy 
\begin{equation}
E_\mathrm{def}=
\frac{\rho}{2}\int d\br d\br^\prime 
\bu(\br,t)(\hat{W}^{\mathrm{ac}})_{\vec{r}-\vec{r}'}^2 \bu(\br^\prime,t)    
\end{equation}
 of elastic deformations of the crystal where $\rho$ is the its mass density (see SM~\cite{SM} for an explicit derivation of the dynamical matrix). The equations Eq.~\ref{PhononsAcoustic1} and b can also be written as the eigenvalue problem
 $\omega\psi_\mathrm{ac}=\hat{\mathcal{H}}_{\mathrm{ac}}(\vec{q})\psi_\mathrm{ac}$ with the 6-component vector $\psi_{\mathrm{ac}}=\{\vec{p}, \hat{W}_\vec{q}^{\mathrm{ac}} \vec{u} \}$ 
 and the matrix $\hat{\mathcal{H}}_{\mathrm{ac}}(\vec{q})$ given by
\begin{equation}
\label{HamiltonianPhonons}
\hat{\mathcal{H}}_\mathrm{ac}(\vec{q})=\begin{pmatrix}
0 & -i \hat{W}_{\vec{q}}^{\mathrm{ac}} \\
i \hat{W}_{\vec{q}}^{\mathrm{ac}} & 0
\end{pmatrix}.
\end{equation}


The interactions between helicons and acoustic phonons come from relativistic effects (``inductive coupling''~\cite{HeliconReview1}).
On the one hand, the dynamics of the crystalline lattice lead to the modification
\begin{align}
    \bE(\br,t)\rightarrow \bE(\br,t)
    +\frac{1}{c}\left[\partial_t \bu(\br,t)\times \bB\right]
    \label{Emodification}
\end{align}
of the electric field $\bE$ in the reference frame of the displaced atoms, where, in the long-wave limit under consideration, the displacement $\bu(\br,t)$ 
varies smoothly compared to atomic length scales.
The currents $\bj(\br,t)$ in the helicon waves also lead the forces 
\begin{align}
    \vecf(\br,t)=\frac{1}{c}\left[\bj (\br,t)\times\bB\right]
    \label{ForceElasticDeform}
\end{align}
(per unit volume) acting on the lattice of the crystal and thus contributing to the displacements $\bu(\br,t)$. Equations \eqref{Emodification} and \eqref{ForceElasticDeform} describe the interactions between the electromagnetic field and the deformations of the lattice and thus account for the interactions between helicons and phonons. 

The dynamics of coupled helicons and phonons  is conveniently described by the $15$-component vector 
$\psi_{\mathrm{pl-ac}}=\{\sqrt{\varepsilon_\infty}\vec{E}, \vec{H}, 4\pi\vec{j}/\omega_{\mathrm{p}}\sqrt{\varepsilon_\infty}, \sqrt{4\pi \rho} \vec{p}, \sqrt{4\pi \rho} \hat{W}_\vec{q}^{\mathrm{ac}} \vec{u}  \}^{\mathrm{T}}$,
where the first three vectors describe helicons [cf. Eqs.~\eqref{Continuity}-\eqref{Maxwell2}] and the last two vectors describe the phonon degrees of freedom. The collective dispersion of coupled helicons and phonons is given by the solution of the eigenvalue problem 
$\omega \psi_{\mathrm{pl-ac}}(\vec{q})=\hat{\mathcal{H}}_{\mathrm{pl-ac}}(\vec{q}) \psi_{\mathrm{pl-ac}}(\vec{q})$
with the ``Hamiltonian'' $\hat{\mathcal{H}}_{\mathrm{pl-ac}}$ given by
\begin{align}
\label{HamiltonianFullacoustic}
\scalemath{0.98}{\hat{\mathcal{H}}_{\mathrm{pl-ac}}=\begin{pmatrix}
0 & -\bar{c} \hat{K}_{\vec{q}} & - i \omega_\mathrm{p} \hat{1} & 0 & 0\\
\bar{c} \hat{K}_{\vec{q}} & 0 & 0 & 0& 0\\
 i \omega_\mathrm{p} \hat{1} & 0 & -i \omega_{\mathrm{c}} \hat{L} & i \gamma_\mathrm{ac} \hat{L} & 0  \\ 0&0& i \gamma_\mathrm{ac} \hat{L} & 0  & -i \hat{W}_{\vec{q}}^{\mathrm{ac}} \\ 0& 0& 0 & i \hat{W}_{\vec{q}}^{\mathrm{ac}} & 0 
\end{pmatrix}},
\end{align}
where the frequency $\gamma_\mathrm{ac}=\omega_\mathrm{c}\sqrt{n m/\rho}$ describes the strength of the coupling between helicons and acoustic phonons.

The structure of the matrix~\eqref{HamiltonianFullacoustic}
can be understood as follows. The top left $3\times 3$ block (whose each element is a $3\times 3$ matrix) matches the matrix~\eqref{HamiltonianPlasma} that describes the helicon degrees of freedom in the absence of phonons. The bottom right $2\times 2$ block describes the dynamics of phonons and has eigenvectors that correspond to the harmonic modes of the crystal. The other terms describe the helicon-phonon coupling and are obtained from Eqs.~\eqref{Emodification} and \eqref{ForceElasticDeform} taking into account that $\partial_t \bu(\bq,t)=\bp(\bq,t)$.

The dispersions of the hybridized helicon-phonon excitations, given by the eigenvalues of the matrix~\eqref{HamiltonianFullacoustic}, are plotted in Figs.~\ref{fig:AllDispersions}a-c for a crystal of potassium 
in the magnetic field $B=20\;\hbox{T}$. The character of helicon-phonon hybridisation is qualitatively different for different branches of the dispersion of acoustic phonons.
Helicons do not interact with the antisymmetric combination of two transverse phonon modes. The interaction between helicons and the symmetric combination of the transverse phonon modes
leads to a gap in the dispersion of the resulting excitations at all wave vectors.

By contrast, the interaction of longitudinal phonons with helicons vanishes for $\bq\parallel\bB$ 
and is finite for other wave vectors.
Their hybridisation leads to the formation of two dispersion bands that intersect at the wave vectors
$\bq_\text{Weyl}^\pm=\pm {\bf e}_z s_l/\alpha$, at which the frequencies of phonons and helicons match ($s_l|\bq|=|\alpha q_z \bq|$) and which are parallel to the magnetic field $\bB$. The interactions lead to the splitting between the bands at all other wave vectors.  
Therefore, the interaction between helicons and the longitudinal phonons leads to the formation of Weyl excitations near the wave vectors $\bq_\text{Weyl}^\pm$.

Near the Weyl points, the Hamiltonian of the Weyl excitations as a function of the wave vector $\bk=\bq-\bq_\text{Weyl}^\pm$ measured from the Weyl points is given by
\begin{align}
\label{HeliconPhonon2}
\hat{\mathcal{H}}_{\mathrm{w}}^\pm
(\vec{k})=
\frac{s_l}{\alpha}+
\frac{3}{2}s_l k_z\pm
\begin{pmatrix} 
\frac{1}{2} s_l k_z  & i s_\perp^{\mathrm{ac}} k_-  
\\  - i s_\perp^\mathrm{ac} k_+ & - \frac{1}{2}s_l k_z \end{pmatrix},    
\end{align}
where $k_{\pm}=k_x\pm i k_y$ and the matrix acts in the space of the phonon and helicon bands that are being hybridized. The transverse velocity of the Weyl excitations 
$s_\perp^{\mathrm{ac}}=\gamma_\mathrm{ac} \bar{c}/2\omega_\mathrm{p}$ is determined by the strength of the helicon-phonons interactions. 
The Weyl nodes near the wave vectors $\bq_\text{Weyl}^\pm$
have topological Chern numbers~\cite{Burkov:Annals2018}
$C^\pm=\frac{1}{2\pi}\int_S \bnabla_\bk\times \bra{\psi_+(\bk)}i\bnabla_\bk
\ket{\psi_+(\bk)}\, d\bS=\mp 1$, where $\psi_+(\bk)$ is the eigenvector 
of the Hamiltonian~\eqref{HeliconPhonon2} that describes the quasiparticle state
with the wavevector $\vec{k}$
and $S$
is an arbitrary closed surface 
in reciprocal space surrounding the Weyl node.

\emph{\color{blue} Interaction between helicons and optical phonons.} 
In certain materials, such as polar crystals or ionic semimetals, the elementary cell consists of positively and negatively charged ions, and electric polarization may be caused by a relative displacement 
of the atoms within the cell. 
This leads to strong interactions of helicons with optical phonons, which, unlike the case of acoustic phonons, are non-relativistic in nature.
Such interactions lead to the emergence of additional pairs of Weyl excitations, distinct from those generated by the interactions of helicons with acoustic phonons. 


In what follows,
we consider a crystal with two ions in the elementary cell and introduce the relative displacement $\bw(\br,t)$ between the ions. 
Our results can be generalized to the case of an arbitrary number of atoms in the elementary cell.

Due to a change in the electric polarization, the displacement $\bw$ affects the electric induction
\begin{align}
    \vec{D}(\vec{r},t) =\varepsilon_\infty \bE(\vec{r},t) + 4\pi \chi \bw (\vec{r},t)
    \label{DmodifiedByW}
\end{align}
in the crystal,
where $\varepsilon_\infty$ is determined by the polarization of the electron shells of the atoms in an electric field, and the quantity $\chi$ describes the response of the polarization to $\bw$~\footnote{For polar crystals, $\chi=\omega_{l} \sqrt{(\epsilon_0-\epsilon_\infty)/4\pi}$ is usually parametrized in terms of the static dielectric constant $\epsilon_0$ and its value $\epsilon_\infty$ in the high-frequency limit}.
In the absence of helicons, the long-wave dynamics of displacement $\bw(\br,t)$ are governed by the equation of motion~\cite{HuangOriginal}
\begin{equation}
\label{PhononsOptical1}
\partial_t^2 \vec{w} (\vec{q},t)=-(\hat{W}_\vec{q}^{\mathrm{op}})^2 \vec{w} (\vec{q},t) + \chi \vec{E}(\vec{q},t),
\end{equation}
where the dynamical matrix 
$\hat{W}^\mathrm{op}_{\vec{q}}$
determines the dispersion of optical phonons (see Supplemental Material~\cite{SM} for a specific form of $\hat{W}^\mathrm{op}_{\vec{q}}$) in the absence of phonon interaction with the electromagnetic fields.

Similarly to the case of 
acoustic phonons, the collective modes of helicons interacting with optical phonons can be obtained as a solution of the eigenvalues problem
$\hat{\mathcal{H}}_{\mathrm{pl}-\mathrm{op}}\psi_{\mathrm{pl-op}}=\omega \psi_{\mathrm{pl-op}}$, where $\psi_{\mathrm{pl-op}}=\{\sqrt{\varepsilon_\infty}\vec{E}, \vec{H}, 4\pi\vec{j}/\omega_{\mathrm{p}}\sqrt{\varepsilon_\infty}, \sqrt{4\pi} \vec{p}, \sqrt{4\pi } \hat{W}_\vec{q}^{\mathrm{op}} \vec{u}  \}^{\mathrm{T}}$. 
The resulting effective Hamiltonian $H_\mathrm{pl-op}$ is given by
\begin{align}
\label{HamiltonianFulloptical}
\scalemath{0.96}{\hat{\mathcal{H}}_{\mathrm{pl}-\mathrm{op}}=\begin{pmatrix}
0 & -c \hat{K}_{\vec{q}} & - i \omega_\mathrm{p} \hat{1} & - i \gamma_{\mathrm{op}} \hat{1} & 0 \\
c \hat{K}_{\vec{q}} & 0 & 0 & 0& 0\\
 i \omega_\mathrm{p} \hat{1} & 0 & i \omega_{\mathrm{c}} \hat{L} & 0 & 0  \\ i \gamma_{\mathrm{op}} \hat{1} &0& 0 & 0 & - i\hat{W}^\mathrm{op}_{\vec{q}} \\ 0 & 0& 0 & i\hat{W}^\mathrm{op}_{\vec{q}} & 0
\end{pmatrix}},
\end{align}
where $\gamma_{\mathrm{op}}=\chi \sqrt{4\pi/\epsilon_\infty}$ describes the strength of the coupling between helicons and optical phonons. The off-diagonal blocks in the matrix $\hat{\mathcal{H}}_{\mathrm{pl-ac}}$ and $\hat{\mathcal{H}}_{\mathrm{pl-op}}$, given by Eqs.~(\ref{HamiltonianFullacoustic}) and (\ref{HamiltonianFulloptical}), have different structure that reflects the distinct physical mechanisms of helicon interactions with acoustic and optical phonons.

In what follows, we neglect the dispersion of longitudinal  and transverse optical
phonons, i.e. assume that their frequencies $\omega_l$ and
$\omega_t$ are momentum-independent, which provides a good approximation for realistic crystals.
Also, we assume that the frequencies 
$\omega_l$ and $\omega_t$ are of the same order of magnitude but differ. These assumptions are satisfied for semiconductors with the anisotropic wurtzite crystalline structure (e.g. CdSe, AlN, InN, etc)~\footnote{We also assume that the external magnetic field is directed along the c-axis of the crystal} or in semiconductors with the zincblend  structure (e.g. InAs, InSb, etc) in the presence of a uniaxial strain. We also assume that the semiconductor crystal is heavily doped $\omega_\mathrm{l},\omega_\mathrm{c}\ll\omega_\mathrm{p}$ that implies dispersion curves for both helicons and optical phonons are well separated from the ones for high-frequency plasma Langmuir, R- and L- waves~\footnote{We note that the emergence of the Weyl helicon-phonon waves does not rely on this assumption. Instead, it ensures that the renormalization of phonon dispersion curves due to their interactions with the three gapped high-frequency magnetoplasma modes can be safely neglected and simplifies the analysis of the dispersion relations for the mixed excitations.}  

The dispersion relations of excitations obtained as the eigenvalues of the matrix 
$\hat{\mathcal{H}}_{\mathrm{pl-op}}$ are shown in  
Fig.~\ref{fig:AllDispersions}d-f,
with the numerical values of frequencies and wave vectors given for heavily doped CdSe for the magnetic field $B=45\; \hbox{T}$ and electron concentration $n=4\cdot 10^{19}\; \hbox{cm}^{-3}$ (see Supplemental Material~\cite{SM} for the details of the estimates).

Similarly to the case of acoustic phonons, the dispersions of non-interacting helicons and longitudinal optical phonons intersect along 2D lines, which for optical phonons are given by $\omega_l/\alpha=|q_z q|$. Interactions split these intersecting branches at all wave vectors other than
$\bq^\pm_\text{Weyl}
=\pm {\bf e_z}\sqrt{\omega_l/\alpha}$
parallel to the magnetic field.
Indeed, for this direction of wave vector, the electric field of helicon waves is orthogonal to the polarisation and displacement of longitudinal phonons, and the helicon-phonon interaction vanishes.
As a result, the hybridised helicon-phonon excitations have two bands that touch at the Weyl points at the wave vectors $\bq^\pm_\text{Weyl}$.
As a function of small deviations $\bk=\bq-\bq^\pm_\text{Weyl}$ of momentum from Weyl points, the Hamiltonian of the Weyl excitations is given by
\begin{equation}
\label{HeliconPhonon3}
\hat{\mathcal{H}}_{\mathrm{w}}^\pm
(\vec{k})=
	\omega_{\mathrm{l}}+
	\sqrt{\alpha \omega_{\mathrm{l}}} k_z\pm
	\begin{pmatrix} 
		\sqrt{\alpha \omega_{\mathrm{l}}} k_z  & i s_\perp^{\mathrm{op}} k_-  
		\\  - i s_\perp^\mathrm{op} k_+ & - \sqrt{\alpha \omega_{\mathrm{l}}} k_z \end{pmatrix},    
\end{equation}
where $s_\perp^{\mathrm{op}}= \gamma_{\mathrm{op}} \omega_\mathrm{c} c/2 \omega_{\mathrm{p}}^2 $ is the transverse velocity of Weyl excitations.
The Weyl nodes near wave vectors $\bq_\text{Weyl}^\pm$ have
topological Chern numbers $C^\pm=\mp 1$.


\emph{\color{blue} Discussion.}
A prominent manifestation of the Weyl spectrum topology is a possible presence of protected surface arc states~\cite{Burkov:Annals2018,ZHasan:reviewDiscovery,WanVishwanath:WeylFirst,Armitage:WeylReview}. 
The existence of such arcs on the surface of a crystal requires that the excitations have a global gap in the reciprocal-space plane perpendicular to the surface.  Provided that the global gap is present, the presence of the arc states is dictated by the bulk-edge correspondence~\cite{Armitage:WeylReview}.

In the case of acoustic phonos, the derived above Weyl excitations do not have such a gap and, therefore, lack arc surface states. By contrast, the dispersion of the mixed modes involving optical phonons has the global gap for any reciprocal-space section between the Weyl nodes. As for the other sections, the global gap naturally appears if the optical phonons are dispersive (that is not captured by the hydrodynamic scheme we employ in this Letter). We can get insights into the behavior of the arc states if we consider the evolution of the mixed modes spectra with the longitudinal optical phonons frequency $\omega_\mathrm{l}$ treated as a free parameter. If $\omega_\mathrm{l}$ exceeds $\omega_\mathrm{c}$, longitudinal phonons and helicons do not intersect, and the Weyl nodes do not emerge. 
If $\omega_\mathrm{l}$ crosses $\omega_\mathrm{c}$ and further decreases, the pair of Weyl nodes emerges at the infinity ($q_z\rightarrow \pm \infty$) and then moves towards $q_z=0$. Thus,
the surface arc states connect the projections of Weyl nodes in different Brillouin zones to the 
surface.
The exact shape of the arcs away from their ends is non-universal and depends on the details of the phonon and helicon dispersion in the entire Brillouin zone.

The characteristic wavevectors and frequencies of Weyl excitations are very different in the cases of acoustic and optical phonons.
For acoustic phonons, their mixing with helions has been reported in potassium~\cite{HeliconPhononExpK1,HeliconPhononExpK2,HeliconPhononExpK3}, aluminium~\cite{HeliconPhononExpAl}, indium~\cite{HeliconPhononExpIn}, lead telluride~\cite{HeliconPhononExpPbTe}  and cadmium arsenide~\cite{HeliconPhononExpCdAs}, is prominent within the MHz$\sim$GHz range and can be tuned by an external magnetic field. In the case of optical phonons, the frequencies of the Weyl excitations are determined by the frequencies of optical phonons that typically lie in the THz range. Good material candidates would have: (i) low frequencies of optical phonons and at least a moderate splitting between their longitudinal and transverse modes; (ii) high mobilities
of charge carriers.
Among semiconductors extensively studied in the context of optoelectronics and plasmonics~\cite{MaterialsReview}, these conditions are well satisfied for CdSe and strained InSb.

The discussed phonon-helicon excitations is a rather favourable playground for Weyl physics and observing associated fundamental phenomena.
The predicted Weyl excitations are highly tunable, with the dispersion curves that can be tuned by the external magnetic field,
strain, electric currents, etc. 
The relatively small momenta of these excitations allow for them to be readily excited by electromagnetic pulses in the THz and GHz ranges.
The discussed systems present time-reversal-symmetry-breaking Weyl semimetals~\cite{Armitage:WeylReview} (effectively, Weyl semimetal with only two symmetric Weyl cones, as opposed to the so-called inversion-symmetry-breaking Weyl semimetals, in which the Weyl nodes lie, in general, at different energies). 
Such Weyl semimetals are extremely rare in solid-state materials and have not been observed until recently~\cite{WSExpNew1,WSExpNew2}.

The predicted Weyl excitations and the surface arc states can be probed by phononic techiques (Raman and inelastic X-ray scattering experiments), as well as various techniques designed for plasmon-polaritons, including angle-resolved reflection experiment~\cite{PlasmaTop1,PlasmaTopExp}. The unidirectional
nature of the topological-surface-arc excitations is promising for applications in  the context of topological and non-reciprical plasmonics~\cite{PlasmonicsReview}.

{\it\color{blue} Conclusion and outlook.}
We have shown that a generic conductor in a magnetic field hosts
Weyl excitations, which emerge as a result of the hybridisation of helicon waves and phonons. Such Weyl excitations exist generically due to the interactions of helicons with acoustic phonons. In polar crystals, additional Weyl excitations may emerge due to the interactions of helicons with optical phonons.


In this paper, we considered the emergence of Weyl excitations in topologically trivial materials.
Our approach can be extended to doped topological insulators and Weyl semimetals, in which helicons are impacted by anomalous electronic responses as well as axion electrodynamic effects~\cite{HeliconsWS1,HeliconsWS2}

In this paper, we neglected the effects of dissipation on the considered excitations, assuming quasiparticle and phonon scattering rates to be 
sufficiently smaller than the characteristic frequencies, which corresponds to realistic experimental conditions~\cite{SM}.        
In principle, dissipation effects may not only result in the broadening of Weyl nodes but lead to new intriguing phenomena.
It has been predicted, for example, that in plasmonic and photonic crystals, non-Hermitian perturbations of the effective Hamiltonian can make Weyl nodes evolve into nodal discs~\cite{PlasmaTopNH1} and nodal rings~\cite{PlasmaTopNH2}. We leave the studies of such effects and the conditions necessary for realising them for future studies.

{\it\color{blue} Acknowledgments.}  We acknowledge support from the Australian Research Council Centre of Excellence in Future Low-Energy Electronics Technologies (CE170100039). 

\bibliography{WHRReferences}
\bibliographystyle{apsrev4-1_our_style}


\newpage

\renewcommand{\theequation}{S\arabic{equation}}

\setcounter{section}{0}
\setcounter{equation}{0}
\setcounter{figure}{0}
\setcounter{enumiv}{0}

\clearpage

\setcounter{page}{1}


\begin{center}
	\textbf{\large Supplemental Material for \\
		Weyl excitations via
helicon-phonon mixing in conducting materials}
	\\
	Dmitry K.  Efimkin and Sergey Syzranov
\end{center}
\section{Long-wavelength theory for helicons}
The ``Hamiltonian'' $\hat{\mathcal{H}}_{\mathrm{pl}}(\vec{q})$ given by Eq.~(2a) contains complete full information about the helicon dispersion but is excessive and cumbersome to deal with. To get rid of the high frequency-plasma modes, it is instructive to rewrite $\hat{\mathcal{H}}_{\mathrm{pl}}(\vec{q})$ in the eigenbasis for $\hat{\mathcal{H}}_{\mathrm{pl}}(\vec{q}=\vec{0})$ that admits an analytical solution. For instance, the dispersion of helicons is gapless and therefore vanishes at $\vec{q}=0$, but this is not the case for the high frequency plasma waves. The Langmuir waves is dispersionless and its frequency is $\omega_\mathrm{p}$, while the corresponding frequencies for R- and L-waves are given by
\begin{equation}
\omega_\pm=\sqrt{\omega_\mathrm{p}^2+\left(\frac{\omega_\mathrm{c}}{2}\right)^2}\pm\frac{\omega_\mathrm{c}}{2}.
\end{equation}
The splitting between them is induced by external magnetic field and is equal to the cyclotron frequency $\omega_\mathrm{c}$. 
The basis transformation results in the Hamiltonian $\hat{\mathcal{H}}'_{\mathrm{pl}}(\vec{q})$ that is given by 
\begin{equation}
\hat{\mathcal{H}}'_{\mathrm{pl}}(\vec{q})=
\begin{pmatrix}
\hat{\epsilon}_+ & \hat{\Sigma}_{+0}(\vec{q}) & 0 \\ \hat{\Sigma}_{0+}(\vec{q}) & 0 & \hat{\Sigma}_{0-}(\vec{q}) \\ 0 & \hat{\Sigma}_{-0}(\vec{q}) & \hat{\epsilon}_-
\end{pmatrix}
\end{equation}
Here $\hat{\epsilon}_+=\mathrm{diag}[\{\omega_+, \omega_\mathrm{p}, \omega_-\}]$ and  $\hat{\epsilon}_-=\mathrm{diag}[\{-\omega_-, -\omega_\mathrm{p}, -\omega_+\}]$ are diagonal matrices. Importantly, zero-frequency modes are not coupled directly, but only via transitions to high frequency plasma modes (including both positive and negative frequency sectors). If we reshuffle wave functions for zero-frequency modes as $\psi_{\mathrm{hel}}(\vec{q})=\{(B_x-i B_y)/\sqrt{2}, -B_z,(-B_x-i B_y)/\sqrt{2}\}$, the matrix elements of their coupling with high-frequency plasma waves are given by
\begin{equation*}
\begin{split}
\Sigma_{+0}(\vec{q})&=\begin{pmatrix} i \bar{c} q_z r_-  & \frac{ i c q_+ r_-}{\sqrt{2}}  & 0 \\ \frac{\bar{c} q_-}{2} & 0 & \frac{\bar{c} q_+ }{2} \\ 0 & \frac{i \bar{c} q_- r_+}{\sqrt{2}} & - i \bar{c} q_z r_+ 
\end{pmatrix}, \\
\Sigma_{-0}(\vec{q})&=\begin{pmatrix} - i \bar{c} q_z r_+  & -\frac{i \bar{c} q_+ }{\sqrt{2}}   & 0 \\ \frac{ - c q_-}{2}  & 0 & -\frac{\bar{c} q_+}{2} \\ 0 & - \frac{ i c q_-}{\sqrt{2}}  & i \bar{c} q_z r_-
\end{pmatrix}.
\end{split}
\end{equation*}
Here we have introduced $q_{\pm}=q_x\pm iq_y$,   $r_-=\sqrt{\omega_-/(\omega_++\omega_-)}$ and $r_+=\sqrt{\omega_+/(\omega_++\omega_-)}$.
The other two matrix elements are not independent, but can be obtained by the Hermitian conjugation as
as $\Sigma_{0+}=\Sigma_{+0}^{\dagger}$ and $\Sigma_{0-}=\Sigma_{-0}^{\dagger}$. In the leading order in the small parameter  $\omega_\mathrm{c}/\omega_\mathrm{p}\ll1$,the exclusion of the high-frequency plasma waves is straightforward and results in the following effective low-frequency "Hamiltonian" $\hat{\mathcal{H}}_\mathrm{hel}(\vec{q})$ that is given by  
\begin{equation}
\label{HamiltonianHSM1}
\hat{\mathcal{H}}_{\mathrm{hel}}=-\hat{\Sigma}_{0+}(\vec{q}) \hat{\epsilon}_{+}^{-1}\hat{\Sigma}_{+0} (\vec{q})-\hat{\Sigma}_{0-}(\vec{q}) \hat{\epsilon}_{-}^{-1}\hat{\Sigma}_{-0} (\vec{q}).
\end{equation}
Explicit evaluation results in
\begin{equation}
\label{HamiltonianHSM2}
\hat{\mathcal{H}}_{\mathrm{hel}}(\vec{q})
=\alpha q_z\begin{pmatrix} q_z & \frac{q_x+i q_y }{\sqrt{2}}  & 0 \\ \frac{q_x-i q_y }{\sqrt{2}} & 0 & \frac{q_x+i q_y }{\sqrt{2}} \\ 0 & \frac{q_x-i q_y}{\sqrt{2}} & -q_z
\end{pmatrix} 
\end{equation}
that is presented in the main part of the paper as Eq.~(3). Here $\alpha=\omega_\mathrm{c} \bar{c}^2/\omega_\mathrm{p}^2=B c/4\pi n e$ gives the prefactor in the anisotropic dispersion relation $\omega_{\mathrm{hel}}(\vec{q})=\alpha q_\mathrm{z} q$ for helicons. 

Interestingly, the Hamiltonian  $\hat{\mathcal{H}}_{\mathrm{hel}}(\vec{q})$ given by Eq.~(\ref{HamiltonianHSM2}) can be readily recognized as product of the spin-1 Weyl Hamiltonian and $q_z$-factor.  
The presence of the extra factor not only shapes the anisotropic helicon dispersion, but makes the plane $q_z=0$ in reciprocal space to be special. Indeed, all three eigenvalues of $\hat{H}_{\mathrm{hel}}(\vec{q})$ are degenerate there. As a result, positive and negative branches for the spin-1 Weyl Hamiltonian switch across this plane, and wave function for helicons (in the basis $\psi_\mathrm{hel}=\{B_- , -B_z,B_+\}$) can be presented as
\begin{equation}
\label{WFHelicon}
|\psi_{\mathrm{H}}(\vec{q})\rangle=\begin{pmatrix} \frac{1+|n_z|}{2} (n_x+i n_y) \\ \frac{n_\perp}{\sqrt{2}} \frac{n_z}{|n_z|}\\ \frac{1-|n_z|}{2} (n_x-i n_y)
\end{pmatrix}.    
\end{equation}
Here $\vec{n}=\vec{q}/q$ is the unit vector aligned with the wavevector $\vec{q}$. Generally, propagation of helicons involves both longitudinal and transverse oscillations (of electronic current, as well as electric and magnetic fields). However, if the wavevector $\vec{q}$ and magnetic field are aligned, helicons become circularly polarized transverse wave with no longitudinal component. Their wave function simplifies as   
\begin{equation}
\label{WFHeliconAlligned}
|\psi_{\mathrm{H}}(\vec{q}_z)\rangle=\begin{pmatrix}  n_x+i n_y \\ 0 \\ 0
\end{pmatrix}.    
\end{equation}
It also can be noticed that circular polarization is the same as for helicons propagating along magnetic field as ones propagating in the opposite direction. 

\section{Equations of motion for phonons}
This section discusses the reformulation of equations of motion for phonons, Eqs.~(4a) and (12) in the main text, as an eigenvalue problem. It is instructive to start with acoustic phonons. The corresponding equation of motion, Eq.~(4a), is of the second order in $\partial_t$ and needs to be rewritten as a system of two coupled equations of the first order in $\partial_t$. After the Fourier transform, this can be achieved in a naive way as 
\begin{equation}
\label{PhononsAcoustic1SM}
\begin{split}
\partial_t\vec{u}(\vec{q},t)=\vec{p}(\vec{q},t), \\  
\partial_t\vec{p}(\vec{q},t)=-\hat{D}_{\vec{q}}^{\mathrm{ac}} \vec{u}(\vec{q},t)
\end{split}
\end{equation}
The explicit expression for the dynamical matrix  $\hat{D}_{\vec{q}}^{\mathrm{ac}}$ is given by
\begin{widetext}
\begin{equation}
\hat{D}_{\vec{q}}^{\mathrm{ac}}= \begin{pmatrix} s^2_{\mathrm{l}} q_x^2 + s^2_{\mathrm{t}} (q_y^2+q_z^2)
 & (s_{\mathrm{l}}^2-s_{\mathrm{t}}^2) q_x q_y & (s_{\mathrm{l}}^2-s_{\mathrm{t}}^2) q_x q_z \\ (s_{\mathrm{l}}^2-s_{\mathrm{t}}^2) q_y q_x & s^2_{\mathrm{l}} q_y^2 + s^2_{\mathrm{t}} (q_y^2+q_z^2) & (s_{\mathrm{l}}^2-s_{\mathrm{t}}^2) q_y q_z  \\ (s_{\mathrm{l}}^2-s_{\mathrm{t}}^2) q_z q_x &  (s_{\mathrm{l}}^2-s_{\mathrm{t}}^2) q_z q_y & s^2_{\mathrm{l}} q_z^2 + s^2_{\mathrm{t}} (q_x^2+q_y^2)
\end{pmatrix}.
\end{equation}
\end{widetext}
The resulting system of equations is of the first order in $\partial_t$, but does not have the desirable "Shr\"{o}dinger-like" structure. This can be fixed if we notice that the matrix $\hat{D}_{\vec{q}}^{\mathrm{ac}}=(\hat{W}_\vec{q}^{\mathrm{ac}})^2$ can be presented as a square of another Hermitain matrix $\hat{W}_{\vec{q}}^{\mathrm{ac}}$. This is possible because all eigenvalues for $\hat{D}_{\vec{q}}^{\mathrm{ac}}$ are positive $\{s_\mathrm{l}^2 \vec{q}^2, s_\mathrm{t}^2 \vec{q}^2, s_\mathrm{t}^2 \vec{q}^2\}$. The connection between $\hat{D}_{\vec{q}}^{\mathrm{ac}}$ and $\hat{W}_{\vec{q}}^{\mathrm{ac}}$ can be established as
\begin{equation}
\begin{split}
\label{WMatrixacousticPhononsSM}
\hat{D}_{\vec{q}}^{\mathrm{ac}}=s_\mathrm{l}^2 \vec{q}^2 |\psi_{\mathrm{l}}(\vec{q})\rangle \langle \psi_{\mathrm{l}}(q)| \\ + s_\mathrm{t}^2 \vec{q}^2 |\psi_{\mathrm{t1}}(\vec{q})\rangle \langle \psi_{\mathrm{t1}}(\vec{q})|+s_\mathrm{t}^2 \vec{q}^2 |\psi_{\mathrm{t2}}(\vec{q})\rangle \langle \psi_{\mathrm{t2}}(\vec{q})|, \\
\hat{W}_{\vec{q}}^{\mathrm{ac}}=s_\mathrm{l} q |\psi_{\mathrm{l}}(\vec{q})\rangle \langle \psi_{\mathrm{l}}(\vec{q})| + \\ s_\mathrm{t} q |\psi_{\mathrm{t1}}(\vec{q})\rangle \langle \psi_{\mathrm{t1}}(\vec{q})|+s_\mathrm{t} q |\psi_{\mathrm{t2}}(\vec{q})\rangle \langle \psi_{\mathrm{t2}}(\vec{q})|.
\end{split}
\end{equation}
Here we have introduced eigenstates for longitudinal and trasnverse phonons in the basis $\{u_x, u_y,u_z\}$ as 
\begin{equation*}
\begin{split}
\label{EigenstatesPhononsSM}
|\psi_{\mathrm{l}}(\vec{q})\rangle= \{n_x, &  n_y,    n_z \}^\mathrm{T}, \quad
|\psi_{\mathrm{t1}}(\vec{q})\rangle=\frac{\{ -n_y,  n_x 
, 0 \}^\mathrm{T}}{\sqrt{n_x^2+n_y^2}}, \\
\\|\psi_{\mathrm{t2}}(\vec{q})\rangle&=\frac{\{ -n_x n_z , -n_y n_z, n_x^2+n_y^2  \}^\mathrm{T}}{\sqrt{n_x^2+n_y^2}}.
\end{split}
\end{equation*}
Here $\vec{n}=\vec{q}/q$ is the unit vector aligned with the wavevector $\vec{q}$. The introduced matrix $\hat{W}_{\vec{q}}^{\mathrm{ac}}$ plays the key role in the transformation
\begin{equation}
\psi_{\mathrm{ac}}=\{\vec{p}, \hat{W}_\vec{q}^{\mathrm{ac}} \vec{u} \}
\end{equation}
that recasts the equations of motion for acoustic phonons as an eigenvalue problem with the "Hamiltonian" that is given by  
\begin{equation}
\label{HamiltonianPhononsSM}
\hat{\mathcal{H}}_\mathrm{ac}(\vec{q})=\begin{pmatrix}
0 & -i \hat{W}_{\vec{q}}^{\mathrm{ac}} \\
i \hat{W}_{\vec{q}}^{\mathrm{ac}} & 0
\end{pmatrix}
\end{equation}
and is presented in the main text as Eq.~(6). 

This approach can readily be generalized to the case of optical phonons. The only difference is that the key matrices $\hat{D}_\vec{q}^{\mathrm{ac}}$ and $W_{\vec{q}}^{\mathrm{ac}}$ needs to be substituted by      
\begin{equation}
\begin{split}
\label{WMatrixOpticalPhononsSM}
\hat{D}_\vec{q}^{\mathrm{op}}=\omega_{\mathrm{l}}^2 |\psi_{\mathrm{l}}(\vec{q})\rangle \langle \psi_{\mathrm{l}}(q)| \\ + \omega_{\mathrm{t}}^2 |\psi_{\mathrm{t1}}(\vec{q})\rangle \langle \psi_{\mathrm{t1}}(\vec{q})|+\omega_{\mathrm{t}}^2 |\psi_{\mathrm{t2}}(\vec{q})\rangle \langle \psi_{\mathrm{t2}}(\vec{q})|, \\
\hat{W}_{\vec{q}}^{\mathrm{op}}=\omega_{\mathrm{l}} |\psi_{\mathrm{l}}(\vec{q})\rangle \langle \psi_{\mathrm{l}}(\vec{q})| + \\ \omega_{\mathrm{t}} |\psi_{\mathrm{t1}}(\vec{q})\rangle \langle \psi_{\mathrm{t1}}(\vec{q})|+\omega_{\mathrm{t}} |\psi_{\mathrm{t2}}(\vec{q})\rangle \langle \psi_{\mathrm{t2}}(\vec{q})|.
\end{split}
\end{equation}
Here $\omega_{\mathrm{l}}$ and $\omega_{\mathrm{t}}$ are dispersion relations for the longitudinal and optical phonons. 
\section{Long-wavelength theory for interacting helicons and phonons}
Effective "Hamiltonians" $\hat{\mathcal{H}}_{\mathrm{hel}-\mathrm{ac}}$ and 
$\hat{\mathcal{H}}_{\mathrm{hel}-\mathrm{opt}}$ describing mixing of helicons with acoustic and optical phonons are presented in the main text as Eqs.~(9) and~(13). They contain the complete information on the dispersion relations of the hybrid modes as well as on their nontrivial topology, but are excessive and cumbersome to deal with. It is instructive to truncate the  transverse phonons as well as perturbatively exclude the high-frequency plasma waves. For both considered scenarios, the functional form of the resulting effective "Hamiltonian" is the same 
\begin{equation}
\label{HeliconPhonon1}
\hat{\mathcal{H}}_\mathrm{hel-ph}(\vec{q})=\begin{pmatrix} \omega_\mathrm{l}(\vec{q})  & M^{+h}_{\vec{q}} & 0 \\ M^{h+}_{\vec{q}}  & \hat{\mathcal{H}}_{\mathrm{hel}}(\vec{q}) & M^{h-}_{\vec{q}} \\ 0 & M^{-h}_{\vec{q}} &  - \omega_\mathrm{l}(\vec{q})
\end{pmatrix}.
\end{equation}
The top left and bottom right corner elements correspond to longitudinal phonons (including their negative frequency counterpart states) and $\omega_\mathrm{l}(\vec{q})$ is their dispersion relation. The central block is $3\times 3$ effective Hamiltonian for helicons  $\hat{\mathcal{H}}_\mathrm{hel}(\vec{q})$ given by Eq.~(\ref{HamiltonianHSM2}). The off-diagonal vectors $M^{\pm h}_\vec{q}=\mp i  s_{\perp} q^{-1}\{ q_z q_-, \sqrt{2}q_\perp^2 , - q_z q_+ \}$ describe inter-mode coupling and $M^{h \pm}_{\vec{q}}=(M^{\pm h}_{\vec{q}})^*$. The strength of the coupling between helicons and longitudinal phonons is parameterized by the velocity $s_\perp$ and its explicit expressions for the case of acoustical and optical phonons are given by $s_\perp^{\mathrm{ac}}=\gamma_\mathrm{ac} \bar{c}/2\omega_\mathrm{p}$ and $s_\perp^{\mathrm{op}}= \gamma_{\mathrm{op}} \omega_\mathrm{c} \bar{c}/2 \omega_{\mathrm{p}}^2 $. In the vicinity of the Weyl nodes, this "Hamiltonian" $\hat{\mathcal{H}}_\mathrm{hel-ph}(\vec{q})$ further simplifies into the genuine Weyl Hamiltonian presented in the main text as Eqs.~(10) and (14). 


\section{Parameters for estimates}

\subsection{Acoustic phonons}
Mixing between acoustic phonons and helicons has been reported in potassium~\cite{HeliconPhononExpK1,HeliconPhononExpK2,HeliconPhononExpK3}, aluminium~\cite{HeliconPhononExpAl}, indium~\cite{HeliconPhononExpIn}, lead telluride~\cite{HeliconPhononExpPbTe}  and cadmium arsenide~\cite{HeliconPhononExpCdAs}. We have used the set of parameters that corresponds to potassium and is given by~\cite{HeliconPhononExpK3}: electron density $n=2.65 \; 10^{22} \; \hbox{cm}^{-3}$, effective mass $m=1.2 m_0$ where $m_0$ is free electron mass, optical dielectric constant $\epsilon_\infty\approx 1$, speed of longitudinal $s_\mathrm{l}=2.0\times 10^5 \; \hbox{cm}/\hbox{s}$ and transverse $s_{\mathrm{t}}= 7.4 \times 10^4 \; \hbox{cm}/\hbox{s}$ phonons, and electron scattering time  $\tau_\mathrm{e} \approx 0.4\times 10^{-10}\; \hbox{s}$ (at  $T=1.2\; \hbox{K}$)\cite{Libchaber:HeliconDamping}. We have also chosen the external magnetic field  
$B=20\;\hbox{T}$. The resulting plasma $\omega_\mathrm{p}\approx 8.3~\hbox{PHz}$ and cyclotron $\omega_\mathrm{c}\approx 2.8~\hbox{THz}$ frequencies ($\omega_\mathrm{p}\approx 5.5~\hbox{eV}$ and $\omega_\mathrm{c}\approx 1.9~\hbox{meV}$) are separated by a few orders of magnitude.   

The resulting strength of the coupling between helicons and acoustic phonons is parameterized by the frequency $\gamma_{\mathrm{ac}}\approx16\; \hbox{GHz}$. The resulting wave vector and frequency for the Weyl point are $q_\mathrm{Weyl}^\pm=0.53\;\mu m^{-1}$ and $\omega_{\mathrm{Weyl}}=1.1\; \hbox{GHz}$. The transverse velocity across the Weyl node is $s_{\perp}^{\mathrm{ac}}=2.9 \; 10^4\; \hbox{cm}/\hbox{s}$. 

The quality factor for helicons (the ratio of their decay rate to frequency~\cite{Libchaber:HeliconDamping}) is given by $1/\omega_\mathrm{c} \tau_\mathrm{e}\approx 10^{-3}$.  This demonstrates that the effect of dissipation at the Weyl nodes is very small and they are readily observable even at elevated temperatures.


\subsection{Optical phonons}
Frequencies of optical phonons in typical polar semiconductors lie in the range $5-20\; \hbox{THz}$~\cite{MaterialsReview, OpticalPhononsReview2}. In this paper, we consider $\hbox{CdSe}$, a material characterized by relatively small phonon frequencies, $\omega_\mathrm{l}\approx6\;\hbox{THz}$ and $\omega_\mathrm{t}\approx5.3\; \hbox{THz}$, as well as a relatively small effective electron mass $m\approx 0.12 m_0$, thus requiring lower magnetic fields (cyclotron frequencies) to achieve intersection of optical-phonon and helicon dispersion curves (Another possible candidate is strained InSb). The other parameters are~\cite{OpticalPhononsCdSe1}: static dielectric constant $\epsilon_0\approx10.2$, its value $\epsilon_\infty\approx 6.3$ in the high frequency limit. We also consider electron density $n\approx 10^{19}~\hbox{cm}^{-2}$ and the magnetic field and the magnetic field  
$B=25\;\hbox{T}$. The corresponding electron scattering time can be estimated as $\tau_\mathrm{e} \approx 3\times 10^{-13}\; \hbox{s}$ at $T=80~\hbox{K}$~\cite{Madelung:CrystalData}. The resulting plasma $\omega_\mathrm{p}=186~\hbox{Thz}$ and cyclotron $\omega_\mathrm{c}=36~\hbox{THz}$ frequencies ($\omega_\mathrm{p}\approx 124~\hbox{meV}$ and $\omega_\mathrm{c}\approx 24~\hbox{meV}$) are reasonably well separated and exceed the frequencies of optical phonons.

The resulting strength of the coupling between helicons and optical  phonons is parameterized by the frequency is $\gamma_{\mathrm{op}}\approx3.7\; \hbox{THz}$. The frequency of the Weyl nodes is pinned to the frequency of longitudinal optical mode, whereas the corresponding wave vector is given by $q_\mathrm{Weyl}^\pm=0.69\;\mu m^{-1}$. The transverse velocity across the Weyl nodes is $s_\perp^{\mathrm{op}}\approx 2.3 \times 10^6\; \hbox{cm}/\hbox{s}$. 

The quality factor for helicons (the ratio of their decay rate to frequency) is given by $1/\omega_\mathrm{c} \tau_\mathrm{e}\approx 10$ and can be further improved by cooling the crystal to helium temperatures or by reducing the doping level (As we discuss in the main text, the condition  $\omega_\mathrm{l},\omega_\mathrm{c}\ll\omega_\mathrm{p}$ is not a requirement for the Weyl helicon-phonon waves, but simplifies analysis of their dispersion relations).
The predicted Weyl nodes can, therefore, be clearly resolved in experiments.



\end{document}